\documentclass[traditabstract]{aa} 
\usepackage{graphicx}
\usepackage{natbib}
\usepackage{txfonts}
\usepackage{lscape}
\usepackage{rotating}
\input psfig.sty
\input epsf.sty
\usepackage{epsfig}

\def\mv{\hbox{M$_{\rm V}$}}

\def\msun{\hbox{M$_\odot$}}
\def\mstar{\hbox{M$\star$}}
\def\fmid{\hbox{F$_{\rm MID}$}}
\def\t4{\hbox{t$_{\rm 4}$}}

\def\cm3{\hbox{cm$^{-3}$}}

\begin{document}
   \title{Spectroscopic Constraints on the Form of the Stellar Cluster Mass Function}


   \author{N. Bastian\inst{1}
       \and I.S. Konstantopoulos\inst{2}
       \and G. Trancho\inst{3,4}
       \and D.R. Weisz\inst{5}
       \and S.S. Larsen\inst{6}
       \and M. Fouesneau\inst{5}
       \and C. B. Kaschinski\inst{7}
       \and M. Gieles\inst{8}
   }

   \institute{Excellence Cluster Universe, Boltzmannstr. 2, 85748 Garching, Germany\label{1}\\
       \email{bastian@usm.lmu.de}
       \and
       Department of Astronomy and Astrophysics, The Pennsylvania State University, University Park, PA 16802, USA \label{2}
       \and
       Gemini Observatory, Casilla 603, La Serena, Chile \label{3}
        \and
        Giant Magellan Telescope Organization, 251S. Lake ave., Pasadena 91001,USA \label{4}
        \and
        Department of Astronomy, Box 351580, University of Washington, Seattle, WA 98195, USA\label{5}
 	\and
       Department of Astrophysics, IMAPP, Radboud University Nijmegen, P.O. Box 9010, 6500 GL Nijmegen, The Netherlands \label{6}
        \and
        Institut f\"ur Astronomie und Astrophysik der Universit\"at M\"unchen, Scheinerstra{\ss}e 1, 81679 M\"unchen, Germany\label{9}
       \and
        Institute of Astronomy, University of Cambridge, Madingley Road, Cambridge, CB3 0HA, UK\label{7}
   }
   
   \date{Received Feb. ZZ, 2012; accepted YYY ZZ, 2012}

 \abstract
{This contribution addresses the question of whether the initial cluster mass function (ICMF) has a fundamental limit (or truncation) at high masses. The shape of the ICMF at high masses can be studied using the most massive young ($<$10~Myr) clusters, however this has proven difficult due to low-number statistics. In this contribution we use an alternative method based on the luminosities of the brightest clusters, combined with their ages. The advantages are that more clusters can be used and that the ICMF leaves a distinct pattern on the global relation between the cluster luminosity and median age within a population.  If a truncation is present, a generic prediction (nearly independent of the cluster disruption law adopted) is that the median age of bright clusters should be younger than that of fainter clusters.  In the case of an non-truncated ICMF, the median age should be independent of cluster luminosity.  Here, we present optical spectroscopy of twelve young stellar clusters in the face-on spiral galaxy NGC~2997.  The spectra are used to estimate the age of each cluster, and the brightness of the clusters is taken from the literature.  The observations are compared with the model expectations of Larsen~(2009) for various ICMF forms and both mass dependent and mass independent cluster disruption.  While there exists some degeneracy between the truncation mass and the amount of mass independent disruption, the observations favour a truncated ICMF.  For low or modest amounts of mass independent disruption, a truncation mass of $5-6 \times10^5$\msun\ is estimated, consistent with previous determinations.  Additionally, we investigate possible truncations in the ICMF in the spiral galaxy M83, the interacting Antennae galaxies, and the collection of spiral and dwarf galaxies present in Larsen~(2009) based on photometric catalogues taken from the literature, and find that all catalogues are consistent with having a truncation in the cluster mass functions. However for the case of the Antennae, we find a truncation mass of a few $\times10^6$ \msun, suggesting a dependence on the environment, as has been previously suggested.}

\titlerunning{Constraining the cluster initial mass function}
\maketitle
%

\section{Introduction}
\label{sec:intro}

Since their discovery in starburst and merging galaxies (e.g., Schweizer 1987, Holtzman et al.~1992), young massive stellar clusters (YMCs, a.k.a. young globular clusters or super star clusters) have been the focus of a plethora of studies (c.f. Portegies Zwart, McMillan, \& Gieles~2010).  These studies have focused on individual clusters (e.g., Smith \& Gallagher 2001; Larsen et al. 2008) as well as full cluster populations (e.g., Konstantopoulos et al.~2009; Adamo et al.~2010) in a variety of galaxy types.   Some characteristics of YMC populations appear to be related to the properties of the host galaxy, with high star-formation rate galaxies (i.e., starbursts) producing more massive clusters (e.g., Larsen~2002, Bastian~2008), and there are indications that the cluster formation efficiency (i.e., the fraction of stars that form in clusters) increases with increasing star-formation rate surface density of the host galaxy (e.g., Meurer et al.~1995; Zepf et al. 1999; Larsen~2004; Goddard et al.~2010; Adamo et al. 2011, Silva-Villa \& Larsen~2011).  Additionally, the form of the cluster luminosity or mass distributions appears to be well described by a power-law with index $\sim-2$, at least on the low mass end (e.g., de Grijs et al. 2003).

Despite these intensive studies, there is still ongoing debate in the literature on some of the basic properties of cluster populations.  Is the mass function of clusters truncated at high masses (e.g., Gieles et al. 2006a,b; Bastian~2008; Larsen~2009; Gieles~2009; Bastian et al. 2012) or does the power-law continue to the highest mass clusters (e.g., Chandar et al. 2010; Whitmore et al.~2010)?  How long do clusters live, and how does disruption affect the cluster population?  Is cluster disruption dependent on the mass of the cluster and its environment?  Or is cluster disruption independent of environment, being dominated by internal processes (see Bastian~2011 for a recent review)?  Here, we focus on the form of the initial mass function of clusters, paying particular attention to whether it is truncated at the high mass end.

Whether the initial mass function of stellar clusters (ICMF) has a truncation at the high mass end is important for a number of reasons. In particular, if there is a truncation, and the truncation varies as a function of environment (e.g., Gieles et al.~2006b; Bastian et al.~2012) this may be reflecting important physics in the star/cluster formation process, or about the distribution of the ISM within galaxies (i.e, size scale of GMCs).  Additionally,  a truncation of the ICMF would have important implications for the globular cluster specific frequency problem (e.g., Kruijssen \& Cooper~2012), with the number of clusters surviving after a Hubble time being strongly dependent on whether a truncation occurs at $\sim10^5$~\msun\ or $\sim10^6$~\msun.  Finally, knowing the form of the ICMF allows for tighter constraints to be placed on the role of disruption in shaping cluster populations (e.g, Gieles \& Bastian~2008; Chandar et al.~2010; Bastian et al.~2012).

The form of the ICMF and the mechanism of cluster disruption leaves a characteristic imprint on the relation between the age and absolute magnitude of ensembles of clusters (Larsen~2009).  For example, if cluster disruption is strong (i.e., the timescale of disruption is short) and independent of environment (a mass-independent dissolution time scale, e.g., all clusters lose 90\% of their mass every decade in age - Fall et al.~2009), then few old ($>50$ Myr) bright clusters are expected to exist within the population.  Additionally, if the cluster mass function is truncated at the upper end  (i.e., the mass function is described by a Schechter function), this also limits the likelihood of finding a bright, old cluster. 

Larsen~(2009) derived the expected relations between the median age vs. absolute magnitude of a cluster population, assuming a range of ICMFs and disruption laws.  He used published catalogues of the brightest and 5th brightest clusters in a sample of spiral and dwarf galaxies, and found evidence that the mass function was indeed truncated at a value of $\sim2 \times10^5$~\msun, and that strong mass independent disruption (90\% per decade in age) scenarios were not favoured.  

In the present work, we apply the Larsen~(2009) analysis to a single galaxy, NGC 2997, and use Gemini spectroscopy to derive the age of the individual clusters, eliminating the age/extinction/metallicity degeneracies known to hamper photometric studies.

\section{Observations}
\label{sec:obs}

NGC~2997 is a face on spiral galaxy, located at a distance of $\sim9.5$~Mpc (distance modulus of 29.9 - de Vaucouleurs~1979).  It hosts a relatively rich young cluster population (Larsen \& Richtler~1999), including one of the brightest YMCs known in a spiral galaxy (M$_{V} \sim -12.9$;  Larsen~2009). The face on orientation of NGC~2997, as well as its relatively large distance, allows us to sample the full galaxy with a single pointing of GMOS (discussed in more detail in \S~\ref{sec:spectra}), mitigating any potential selection biases of focussing on the central regions of the galaxy, giving a more global picture.


\subsection{NGC~2997 cluster sample}
\label{sec:sample}

The sample of clusters was taken from the catalogue of Larsen~(2004), which is based on ground-based U, B, V, H$\alpha$ and I-band imaging.  We refer to Larsen\& Richtler~(1999) and Larsen~(1999) for details of the cluster selection process and derivation of the completeness limits, and only summarise here.  Cluster candidates were selected through {\sc  DAOFIND} routine in IRAF, and a colour cut of $B-V < 0.45$ was applied to remove foreground stars.  Additionally, magnitude cuts were applied in order to remove individual bright stars within NGC~2997, namely \mv\ $< -9.5$ for blue sources ($U-B \leq -0.4 $) and \mv\ $ < -8.5$ for red sources ($U-B > -0.4$).  Artificial star and cluster tests resulted in approximately 80\% completeness at \mv $=-9.5$.  All candidates were visually inspected and large diffuse sources (likely background galaxies or extended high-surface brightness regions within spiral arms) were removed.

Additionally, sources that had significant H$\alpha$ emission associated with them were removed, which has two benefits.  The first is that it is often difficult to distinguish a stellar cluster from an unbound association at young ages (e.g., Larsen~2004; Gieles \& Portegies Zwart~2011; Bastian et al.~2011).  However, once a stellar group has had time to dynamically evolve, it is easier to distinguish between clusters and associations (Gieles \& Portegies Zwart~2011), so restricting to candidates older than $5-7$~Myr helps in removing many spurious sources (i.e., unbound associations).  Secondly, it has been suggested that the rapid removal of gas from a young cluster can disrupt a large fraction of clusters or can lead to significant stellar mass loss within clusters (e.g., Goodwin \& Bastian~2006; Goodwin~2009).  However, within a few crossing times, the clusters should stabilise (e.g., Portegies Zwart et al.~2010), so limiting our sample to older clusters lessens the impact that infant mortality or infant weight loss may have on the population.  

\subsection{Spectroscopy}
\label{sec:spectra}

The brightest 30 candidates were chosen, and a multi-slit spectroscopic mask was made that maximised the number of clusters that could be observed in a single pointing.  When the dispersion axis of two potential targets overlapped, the brighter cluster was chosen.  This resulted in 12 clusters that could be observed in a single pointing.  The absolute V-band magnitudes of these clusters were taken from the Larsen catalogue and the basic properties of the clusters are given in Table~\ref{tab:objects}.

We obtained spectra of 12 sources in NGC~2997 using the Multi-Object Spectroscopy (MOS) mode of the Gemini Multi-Object Spectrograph (GMOS) on Gemini South.  The data were obtained as part of Queue program GS-2009B-Q-29 (PI N. Bastian).  We used the B600 grating and a slit width of 0.75 arcsec, resulting in an instrumental resolution of 110 kms$^{-1}$ at 5000~\AA. The spectroscopic observations were obtained as four individual exposures with an exposure time of 1800 s each (for a total on-target integration time of 7200 s). No atmospheric dispersion corrector (ADC) was available on Gemini South at the time of the observations, causing wavelength-dependent slit losses. We corrected each exposure for this effect using the method of Filippenko (1982).

The basic reductions of the data were done using a combination of the Gemini IRAF package and custom reduction techniques, as described in Appendix~A in Trancho et al.~(2007). 

The age of each cluster was derived using the method presented in Trancho et al.~(2007).  In brief, a template is built for each cluster (avoiding lines affected by nebular emission) using simple stellar population (SSP) models (Gonz{\'a}lez Delgado et al.~2005) and the PPxF method (Cappellari \& Emsellem~2004).  Spectral indices are measured from the template spectra ( $H+He$, $K$, $H8$, H$\gamma_{\rm A}$, $Mgb5177$,$Fe5270$, $Fe5335$;  see Trancho et al. 2007 for the full definitions), along with error estimates, which are then compared to SSP models (measured in the same way) in a least $\chi^2$ sense in order to choose the best fitting model.  The uncertainty in age is derived by running 2000 monte carlo simulations for each cluster, adding random noise to the measured indices, sampled from the estimated errors.  The age uncertainty is then taken as the standard deviation of a log-normal fit to the resulting age distribution.  This method has been shown to be largely independent of the signal-to-noise ratio of the data (Trancho et al.~2007).

\begin{figure}
\includegraphics[width=8.5cm]{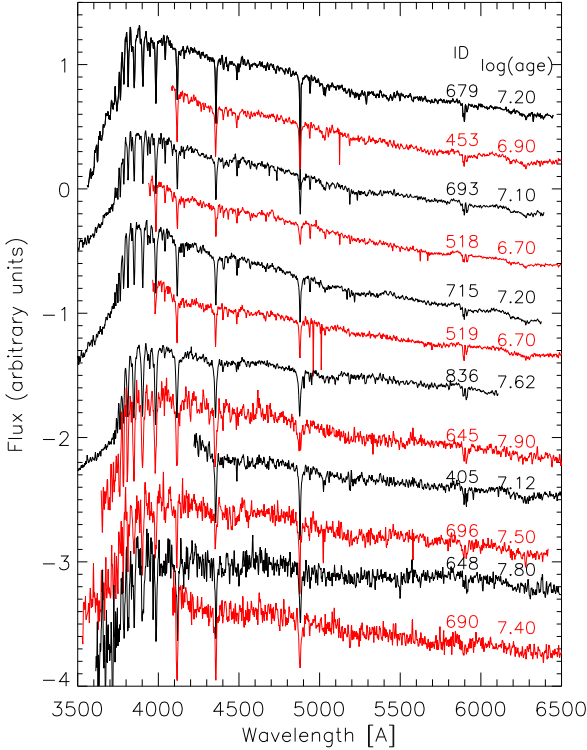}
\caption{The observed spectra of the 12 observed clusters in NGC~2997 presented in this work.  The ID (following L04) and derived spectroscopic (logarithmic) age (in yr) are given for each target.}
\label{fig:spectra}
\end{figure}

The observed spectra are shown in Fig.~\ref{fig:spectra}, along with the estimated age of each cluster.  Additionally, the cluster properties are given in Table~1.  Only the 12 clusters with spectroscopic age determinations are used for determining the cluster mass function.

\begin{figure*}
\includegraphics[width=18cm]{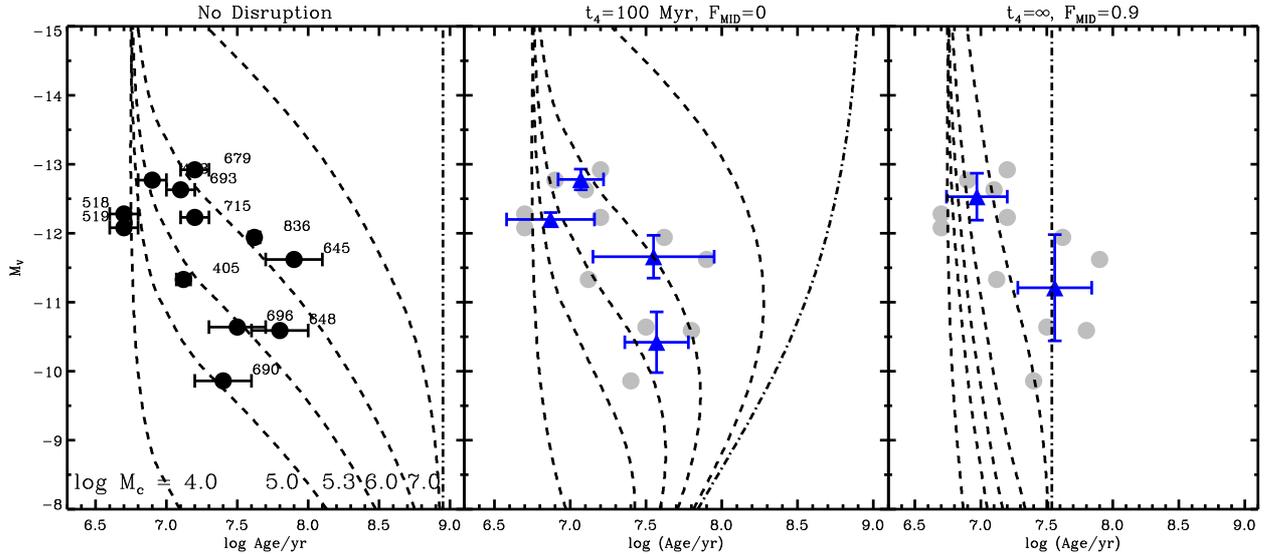}
\caption{Age vs. \mv\ for the twelve clusters in the present study.  In each panel, we also show the expectations of the median \mv, for a given age, following the results of Larsen~(2009).  The model disruption parameters are given at the top of each panel, where \t4\ is the disruption timescale of a $10^4$~\msun\ cluster in the mass dependent cluster disruption scenario, and \fmid\ is the fraction of clusters destroyed every decade in age.  The dashed lines represent models with different values of the Schechter truncation mass, and the dashed-dot lines in each panel represents the case without truncation (i.e., a pure power-law).  In the middle and right panels, the original data are shown in grey, while the binned data (the mean in flux and age, with the error bars representing the standard deviation) are shown as solid (blue) triangles.  Two different bin sizes are shown.}
\label{fig:main}
\end{figure*}

\begin{table*} 
  \begin{tabular}
    {lcccclll} Larsen ID&RA&Dec&Ranking& log (age/yr)&$\sigma$(log age)& M$_{\rm V}$ (mag) & S/N\\
    \hline 
    679&09 45 47.01&$-$31 11 05.3&1& 7.2&0.1&-12.92 & 26\\
    453&09 45 30.98&$-$31 12 43.6&2& 6.9&0.1&-12.77 & 58\\
    693&09 45 48.92&$-$31 10 54.9&3& 7.1&0.1&-12.63 & 35\\
    518&09 45 48.44&$-$31 12 17.2&4& 6.7&0.1&-12.28 & 45\\
    715&09 45 40.63&$-$31 10 50.8&5& 7.2&0.1&-12.23 & 30\\
    519&09 45 32.20&$-$31 12 20.9&7& 6.7 &0.1&-12.08 & 50\\
    836&09 45 33.80&$-$31 10 01.8&8& 7.62&0.05&-11.94 & 25\\
    645&09 45 33.40&$-$31 11 20.9&10&7.9&0.2&-11.62 & 8\\
    405&09 45 35.12&$-$31 13 10.1&11&7.12&0.05&-11.33 & 25\\    
    696&09 45 28.50&$-$31 10 59.8&17&7.5&0.2&-10.64 & 8 \\   
    648&09 45 44.18&$-$31 11 14.6&18&7.8&0.2&-10.59 & 5\\ 
    690&09 45 51.34&$-$31 10 56.3&30&7.4&0.2&-9.86 & 10\\ 
    \hline 
  \end{tabular}
\caption{The ID, location, ranking (in terms of V-band absolute brightness in the population), spectroscopically determined age and error, and absolute V-band magnitude of the clusters in NGC~2997 presented in the current work.}
\label{tab:objects}
\end{table*}

\section{Comparison with Models}
\label{sec:comparison}

The underlying initial cluster mass function (ICMF) and type of disruption strongly affect the relation between the observed age and absolute magnitude of a cluster population.  Larsen~(2009) presented an analytic representation of the $d^2 N/(dtdL)$ distribution, i.e., the number of clusters with an age between $t$ and $t+dt$ and luminosity between $L$ and $L+dL$, that considers functional forms for the ICMF and the mass evolution of the stellar population due to disruption. From this it is possible to derive the median age (as well as the mean and the standard deviation) as a function of M$_{\rm V}$.  He considered various ICMF types (truncated or untruncated) and different disruption laws (mass/environment dependent or independent).  

If disruption does not influence the population, and the mass function has an exponential cutoff (i.e., is described by a Schechter~(1976) type function - $dN/dM_i \sim M_i^{-2} \exp(-M_i/\mstar)$) then the brightest clusters in a population are preferentially young (see also Bastian~2008).  The exact relation between median age and \mv\ depends on the truncation mass, \mstar, with lower \mstar\ values resulting in the brightest clusters being younger.  If cluster disruption is included, in particular if cluster disruption is dependent on the cluster mass, as expected from N-body and analytic considerations (see e.g., Portegies Zwart, McMillan, \& Gieles~2010)  then the older clusters and lower mass clusters are most affected.  This results in a turn-around in the relation between age and \mv\ with fainter clusters also being preferentially young.  Alternatively, if the ICMF does not have a truncation and there is no mass dependent disruption, then the median age of a cluster population should be independent of \mv~ (the vertical dash-dotted line in the first panel of Fig.~\ref{fig:main}).  Including strong mass independent disruption (where a large and constant fraction of the population disappears every decade of age\footnote{There is a subtle difference between all clusters losing a fraction of their mass every decade in age, and a constant fraction of clusters are completely disrupted.  If the mass function is a pure untruncated power-law (with an index of $-2$), then there is no difference in terms of the method used here.  If there is any structure in the ICMF (such as a truncation), in the case of all clusters losing a fraction of their mass, then this structure will shift towards lower masses with time.}) also results in the median age being independent of \mv.  

Three characteristic cases are shown in Fig.~\ref{fig:main} (after Larsen~2009).  In each of the panels six models are shown.  In all models a constant star/cluster formation rate is assumed.  The dashed lines represent models that adopt a Schechter ICMF with \mstar\ values of $1\times10^4$, $1\times10^5$, $2\times10^5$,$1\times10^6$, and $1\times10^7$\msun, from left to right, respectively.  In all cases we adopted an index of $-2$ for the ICMF at the low mass end. The dash-dotted lines represent the results of models that have a pure power-law ICMF (untruncated).  The left panel shows models that have no disruption, while the middle panel shows models with mass dependent disruption (adopting the formalism of Lamers et al.~(2005), with $t_{\rm 4} = 100$~Myr), and the right panel adopts no mass dependent disruption but instead strong mass independent disruption (a long duration infant mortality rate of 90\% - see Whitmore et al.~2007).  The models explicitly include the effects of stochasticity in the cluster population and do not require a fully sampled mass function at young ages.

In all cases, the models represent the {\it median} age of the clusters for a given magnitude.  Hence if the model is a good representation of the data, half of the clusters should lie to the left and half to the right of the curve, at all magnitudes.  From Fig.~\ref{fig:main} it is clear that certain models do not represent the observed data well.  Low or high truncation values (i.e., $\sim10^4$ or $\sim10^7$\msun) are not favoured if there is little or no mass dependent disruption, and low truncation values are also not favoured in the case of strong mass independent disruption.

The data presented in Fig.~\ref{fig:main} show on overall tendency for younger clusters to be brighter than older ones (Spearman's rank correlation coefficient of 0.58, meaning that there is $\sim5\%$ chance of log(age) and M$_{V}$ to show such a relation by chance).  This is a generic prediction for any cluster population that displays a truncation in the upper end of the mass function (e.g., Gieles et al.~2006a, Bastian~2008, Larsen~2009).  If no cluster disruption is assumed, then the data points are most consistent with a truncation mass of a few times $10^5$\msun.  Additionally, while the impact of mass dependent disruption is largely negligible for the observed sample (due to the fact that the brightest clusters tend to be young and massive), strong mass independent disruption models are not favoured by the data.

\subsection{Statistical measures}
\label{sec:statistical}

In order to place the above conclusions on a more statistically robust footing, we have also carried out maximum likelihood tests, comparing the observations to a wide range of models.  For the models, we use the Larsen~(2009) analytic prescriptions, and vary \mstar\ from $1\times10^4$ to 1$\times10^7$\msun, \t4\ from 100 to 10$^{9}$~Myr, and \fmid\ from 0 to 0.9.


To assess the likelihood of our data for a given model, we assume that the uncertainties on both age and $M_{V}$ follow a Gaussian distribution.  From this assumption, we can then write down the likelihood function to be:

\begin{equation}
\mathcal{L} \propto e^\frac{-\chi^2}{2}
\label{eq:likelihood}
\end{equation}

\noindent where 

\begin{equation}
\chi^{2} = \sum_{i=1}^{N} \frac{(\log(age)_{i, obs} - \log(age)_{i, mod})^2} {\sigma_{\log(age)_{i, obs}}^2} + \frac{(M_{V, i, obs} - M_{V, i, mod})^2} {\sigma_{M_{V, i, obs}}^2}
\label{eq:chi}
\end{equation}

\noindent and $N$ is the number of observed data points.  Because we are only interested in relative likelihoods, we do not compute the normalisation term for the likelihood function in Equation \ref{eq:likelihood}.  For this analysis we limit the comparison between observations and models to the range of magnitudes of the observations.

Using this framework, we sampled the likelihood function for 6300 models described in Larsen~(2009).  The ensemble of models that maximise the likelihood function are deemed to represent a set of the most likely model parameters (e.g., truncation mass, \t4, etc).

\subsection{NGC~2997 results}
\label{sec:n2997}

As expected from visual inspection of Fig.~\ref{fig:main}, the results of the maximum likelihood fitting were largely independent of the mass dependent disruption timescale, \t4.  However, for all values of \t4, a truncation in the mass function was favoured by the fitting procedure.  In order to judge the importance of mass independent disruption (i.e., long duration infant mortality) we set the mass dependent disruption timescale to $10^9$~Myr (i.e., essentially infinite).  Doing this, we are left with only two free parameters in the model, \mstar\ and \fmid.

\begin{figure}
\includegraphics[width=9cm]{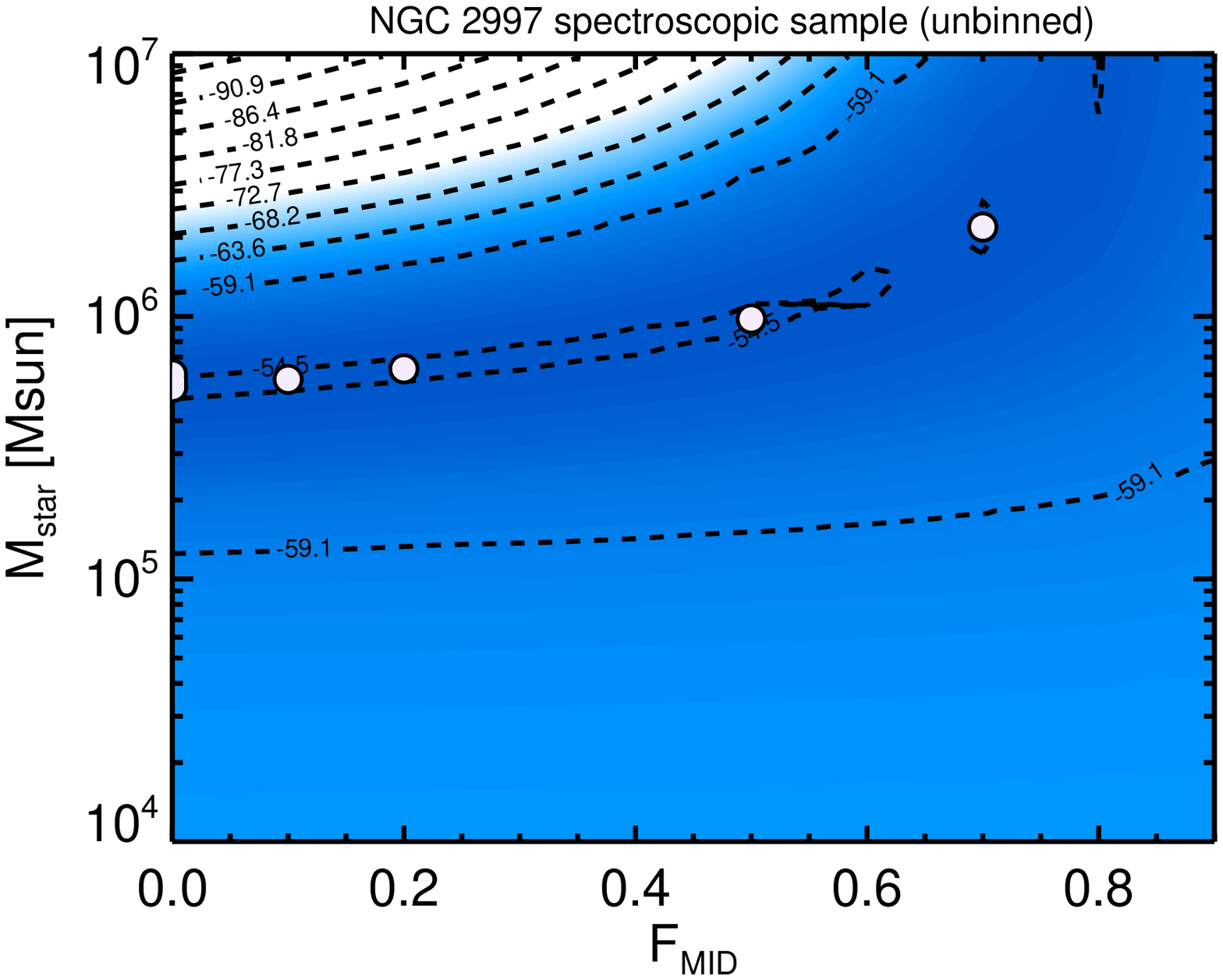}
\includegraphics[width=9cm]{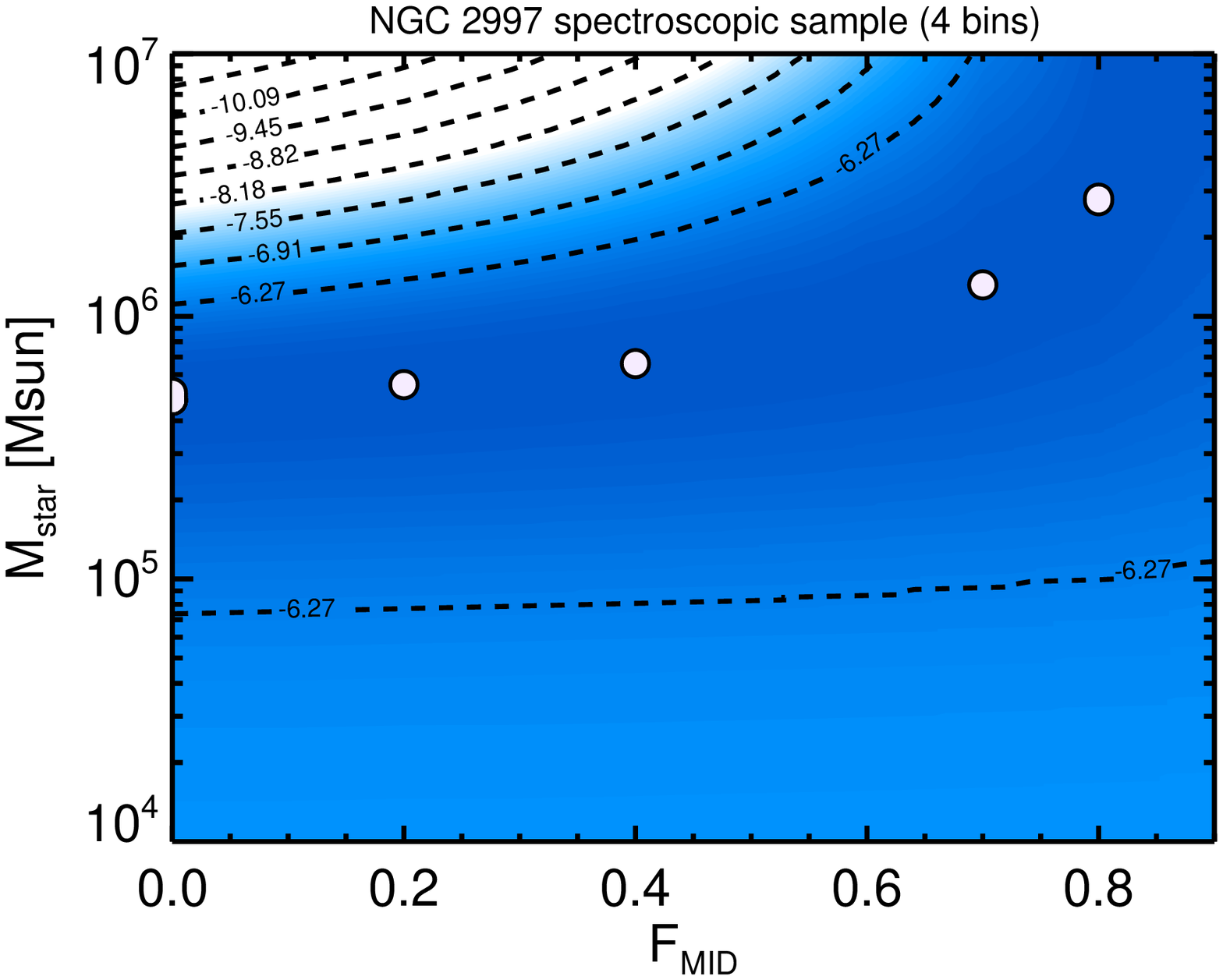}

\caption{Results from the two dimensional statistical analysis (i.e., adopting no mass dependent disruption, and determining the likelihood for the models only varying \mstar\ and \fmid).  The shaded region/contour lines denote the logarithm of the likelihood of the model for the combination of \mstar\ and \fmid, and the filled (white) points denote the maximum likelihood models.  {\bf Top:} The analysis carried out on the unbinned dataset. {\bf Bottom:} The same analysis, but now carried out on the binned dataset (four bins, each with the mean age/flux of three clusters).  The white points from the upper panel denote models that are shown in Fig.~\ref{fig:best_fits}.}
\label{fig:contours}
\end{figure}

\begin{figure}
\includegraphics[width=9cm]{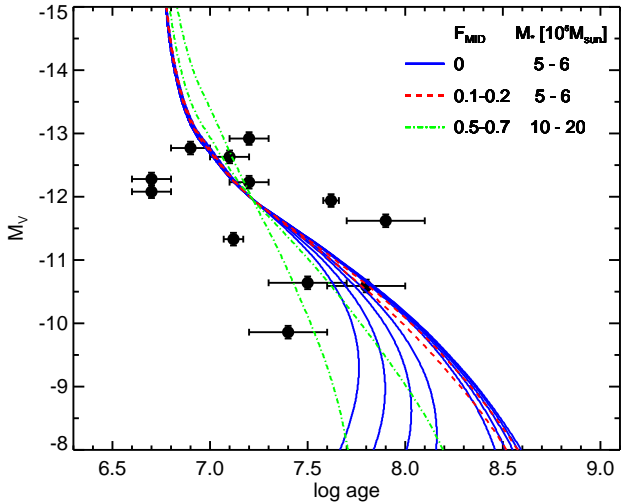}
\caption{The observed age vs. M$_{\rm V}$ relation along with the maximum likelihood models from \S~\ref{sec:n2997}.  The solid (blue curves) represent models that have \fmid=0, and from left to right have \t4\ values ranging from 100~Myr (left) to $10$~Gyr (right).  All these models have \mstar\ values between 5 and 6$\times10^5$\msun.  The dashed (red) lines have similar \mstar\ values, but no mass dependent disruption and a small amount of mass independent disruption.  The dash-dotted (green) lines also represent models with a truncation mass ($\gtrsim10^6$\msun) but have higher mass independent disruption rates.}
\label{fig:best_fits}
\end{figure}

The results from this two dimensional maximum likelihood analysis are shown in Fig.~\ref{fig:contours}.  The darker the shading the higher likelihood (also shown in coarser steps as dashed contour lines).  A clear maximum in this space is seen, a largely horizontal ``valley" with a truncation at $5\times10^5$\msun\ that increases slightly as a function of \fmid.  The solid (white) circles denote the models with the maximum likelihood (i.e., with the same likelihood).  While eleven of the models have \mstar\ between $5-6 \times10^5$\msun\ and \fmid $\leq 0.2$, two of the models had significantly higher \fmid\ values (0.5 and 0.7) and somewhat larger \mstar\ values ($1-2\times10^6$\msun).  These models are shown in Fig.~\ref{fig:best_fits}.

Hence, we find that a truncation in the cluster mass function is favoured independent of the disruption law adopted.  If disruption is dependent on mass and the surrounding environment of a cluster, then a truncation of $\sim5\times10^5$\msun\ is strongly preferred.  A similar conclusion is reached, that a truncation of $\sim5-10 \times 10^5$\msun\ is required,  if the fraction of clusters disrupted every decade in, \fmid, is less or equal to 0.5.  For higher values of \fmid\ a truncation is still preferred, although at higher \mstar\ values.

The largest uncertainty in the value of \mstar\ is due to the disruption law adopted.  If we adopt a mass and environmentally dependent disruption scenario, as found for M83 (Bastian et al.~2011, 2012) and dwarf galaxies (Cook et al.~2012), then the uncertainties are drastically reduced, and instead are due to the limitations of age dating clusters, resulting in an uncertainty in \mstar\ within a factor of two.

\section{Other cluster catalogues}

Additionally, we have run our maximum likelihood method over the data presented in Larsen~(2009).  This catalogue contains the brightest and $5^{\rm th}$ brightest cluster in a photometrically determined sample of cluster populations in spiral and dwarf galaxies.  The fits result in a very similar picture, with \mstar\ slightly lower, $2\times10^5$\msun.

\begin{figure*}
\includegraphics[width=14cm]{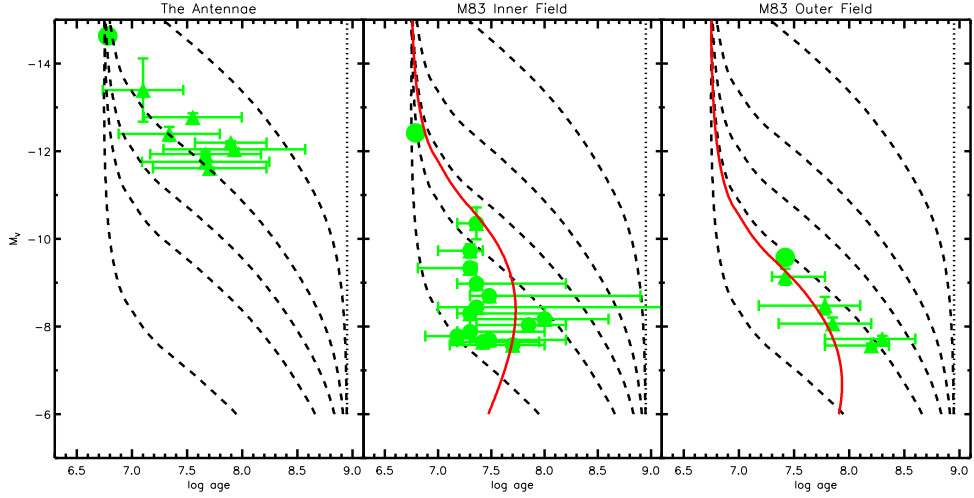}
\caption{Age vs. \mv\ for clusters in the Antennae (left panel - Whitmore et al.~2010), M83 inner region (middle panel - Bastian et al. 2011) and M83 outer region (right panel - Bastian et al. 2011).  Only clusters with ages greater than 6~Myr were used.  In all panels the brightest cluster in the sample is shown as a (green) filled circle.  For the Antennae, the triangles represent the median of the age and \mv\ of clusters in the rest of the catalogue (in bins of four clusters).  For M83, each bin contains 10 clusters.  In the middle/right panels, the model corresponding to the best fit results of Bastian et al.~(2012) are shown as solid (red) lines.  They have been corrected the a Salpeter stellar IMF adopted by the Larsen~(2009) models in \mstar.  For the Antennae data, the error bars represent the standard deviation for each bin.  For the M83 data, the error bars represent the region where 60\% of the data in that bin lie.  We note that the models also show such scatter, i.e. only the median of the models is shown.}
\label{fig:other}
\end{figure*}

Finally, we can look at the photometrically determined properties of cluster populations, taken from the literature.  For this, we use the 50 brightest clusters in the Antennea galaxies as determined by Whitmore et al.~(2010) and the sample of clusters in M83 taken from Bastian et al.~(2011).  For each dataset, we removed sources with ages less than 6~Myr (see \S~\ref{sec:sample} for a discussion on this point) and then looked at the median age and brightness of the population in bins of 4 and 10 clusters in the Antennae and M83, respectively.  The data are shown in Fig.~\ref{fig:other}.  Additionally, for the M83 results, we have overplotted the expected model as derived from the full population study of Bastian et al.~(2012).   These are models with \mstar$=1.6 \times10^{5}$\msun\ and \t4$=130$~Myr for the inner field and \mstar$=5 \times10^{4}$\msun\ and \t4$=600$~Myr for the outer field.

Both galaxies are consistent with having a truncated mass function, with the Antennae having an \mstar\ value of a few $\times10^6$\msun, while \mstar\ is $\sim10^5$\msun in M83.  The value of $\sim10^6$\msun\ found for the Antennae galaxies is similar to that suggested by Zhang \& Fall~(1999) and  Gieles et al.~(2006a) based on the observed luminosity function of the cluster population.  As expected, the cluster mass function extends to higher mass than seen in spiral or dwarf galaxies, however it still appears to be truncated.

\section{Conclusions}
\label{sec:conclusions}

We confirm the findings of Larsen~(2009) that the cluster mass function in spiral galaxies is consistent with a Schechter mass function with a truncation near $2-5 \times10^5$\msun.  Additionally, strong mass independent disruption ($>80-90\%$ per decade in age) is not supported by the current observations.  Comparing the age-\mv\ method to that derived with full population studies in M83 leads to largely consistent results (i.e., that a truncation in the ICMF is needed and that strong mass independent disruption is disfavoured).  Applying this method to the cluster population of the Antennae galaxies shows that it is also consistent with a truncated ICMF, however the truncation mass is significantly higher, a few $\times10^6$\msun.

\begin{acknowledgements}

We would like to thank the referee, Richard de Grijs, for suggestions that help improve the presentation of the paper.  This research was supported by the DFB cluster of excellence 'Origin and Structure of the Universe'.

\end{acknowledgements}


\end{document}